\documentclass[%
 aip,
apl,%
 amsmath,amssymb,
reprint,%
]{revtex4-1}	

\usepackage{color}
\usepackage{graphicx}
\usepackage{dcolumn}
\usepackage{bm}
\usepackage[
colorlinks=true,
linkcolor=blue,
citecolor=blue,
filecolor=blue,
urlcolor=blue]{hyperref}

\begin{document}

\preprint{}

\title[]{Aerogel-based metasurfaces for perfect acoustic energy absorption}

\author{Antonio A. Fern\'andez-Mar\'in}
\altaffiliation{Wave Phenomena Group, Departamento de Ingenier\'ia Electr\'onica, Universitat Polit\`ecnica de Val\`encia, Camino de Vera s/n, 46022 Val\`encia, Spain}
\affiliation{Laboratoire d'Acoustique de l'Universit\'e du Mans, LAUM - UMR 6613 CNRS, Le Mans Universit\'e, Avenue Olivier Messiaen, 72085 LE MANS CEDEX 9, France}

\author{No\'e Jim\'enez}
\altaffiliation{Laboratoire d'Acoustique de l'Universit\'e du Mans, LAUM - UMR 6613 CNRS, Le Mans Universit\'e, Avenue Olivier Messiaen, 72085 LE MANS CEDEX 9, France}
\affiliation{Instituto de Instrumentaci\'on para Imagen Molecular (i3M), CSIC-UPV, Camino de Vera s/n, 46022 Val\`encia, Spain}

\author{Jean-Philippe Groby}
\affiliation{Laboratoire d'Acoustique de l'Universit\'e du Mans, LAUM - UMR 6613 CNRS, Le Mans Universit\'e, Avenue Olivier Messiaen, 72085 LE MANS CEDEX 9, France}

\author{Jos\'e S\'anchez-Dehesa}
\affiliation{Wave Phenomena Group, Departamento de Ingenier\'ia Electr\'onica, Universitat Polit\`ecnica de Val\`encia, Camino de Vera s/n, 46022 Val\`encia, Spain}

\author{Vicente Romero-Garc\'ia}
\affiliation{Laboratoire d'Acoustique de l'Universit\'e du Mans, LAUM - UMR 6613 CNRS, Le Mans Universit\'e, Avenue Olivier Messiaen, 72085 LE MANS CEDEX 9, France}

\begin{abstract}
  The unusual viscoelastic properties of silica aerogel plates are efficiently used to design subwavelength perfect sound absorbers. We theoretically, numerically and experimentally report a perfect absorbing metamaterial panel made of periodically arranged resonant building blocks consisting of a slit loaded by a clamped aerogel plate backed by a closed cavity. The impedance matching condition is analyzed using the Argand diagram of the reflection coefficient, i.e., the trajectory of the reflection coefficient as a function of frequency in the complex plane. The lack or excess of losses in the system can be identified via this Argand diagram in order to achieve the impedance matching condition. The universality of this tool can be further exploited to design more complex metasurfaces for perfect sound absorption, thus allowing the rapid design of novel and efficient absorbing metamaterials.
\end{abstract}

%
\maketitle

Silica aerogels are extremely-lightweight nanoporous materials\cite{Kistler1931}. The frame of these materials consists of an assembly of connected small cross-sections beam-like elements resulting from fused nanoparticles. This particular assembly provides silica aerogel a very low elastic stiffness when compared to rigid silica structure of identical porosity. Aerogels possess a wide variety of exceptional properties such as low thermal conductivity, low dielectric constant, low index of refraction or a very large porosity (80-99.8$\%$) thus providing these materials an extremely low density\cite{Gesser_89}. Because of this large porosity and therefore of the very large available contact area, they have been used as filters, absorbent media or waste containment (see Refs. [\onlinecite{Cooper_89,Gesser_89,Komarneni_93}] and references therein). They have also been applied as catalysts or even to capture cosmic dust \cite{Hrubesh_98, Tsou_95}. Similarly, their low thermal conductivity, which so far seems to be their most interesting property compared to any other elastic or poroelastic material, has been exploited in various commercial applications including thermal insulation \cite{Herrmann_95}, heat and cold storage devices \cite{Hrubesh_98,Fricke_95}.
\par
In acoustics, aerogels are used as impedance matching materials to develop efficient ultrasonic devices \cite{Gronauer_86,Gerlach_92} or sound absorbing materials for anechoic chambers\cite{Hrubesh_98,Gibiat_95}. Beyond these properties, silica aerogel plates are excellent candidates to design new types of membrane metamaterials, since they exhibit subwavelength resonances and present absorption efficiency\cite{Guild_2016,Geslaina_2018}. Effectively, the use of membrane metamaterials to control acoustic waves has shown an increasing interest in recent years\cite{huang2016membrane}. Membrane and plate metamaterials have been employed in the past to design efficient absorbers\cite{mei2012dark,Ma_2014,yang2015subwavelength}, e.g., using a single membrane backed by a cavity \cite{Ma_2014,Romero_2016}, which can present deeper subwavelength resonances as compared with absorbing metamaterials based in air cavities\cite{li2016acoustic,jimenez2016,yang2017optimal,peng2018composite}. Moreover, double negative acoustic metamaterials\cite{yang2013coupled} can be achieved by combining a lattice of membranes, which provides a negative effective mass density \cite{Yang_2008}, with subwavelength side-branch resonators, which provides a negative effective bulk modulus\cite{Lee_2010}. In addition, periodic arrangements of clamped plates have been efficiently used to control harmonic generation\cite{Zhang_2016} or solitary waves in the nonlinear acoustic regime\cite{Zhang_2017}. 
\par
\begin{figure}[t]
	\centering
	\includegraphics[width=\columnwidth]{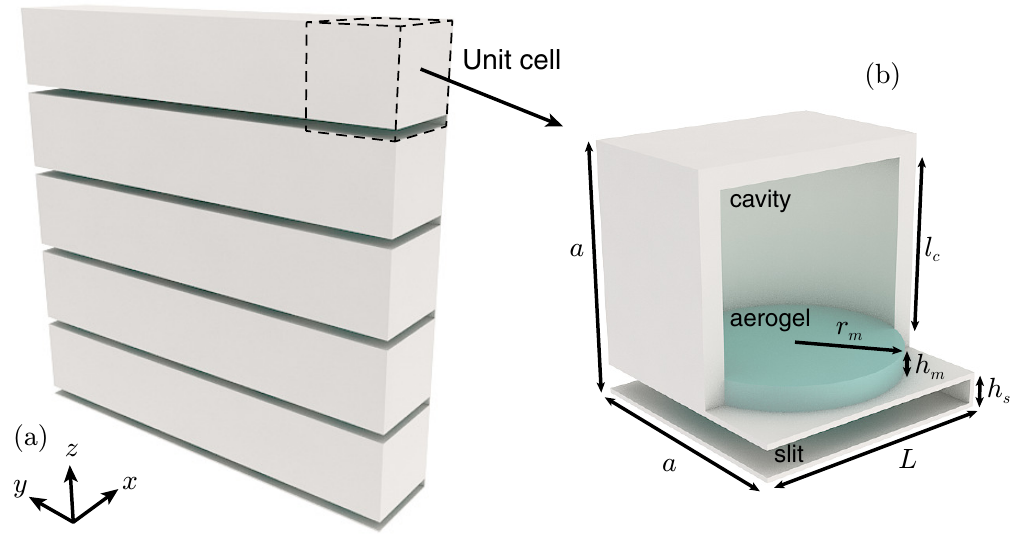}
	\caption{(color online) (a) Scheme of the panel under consideration constructed by a slit, the aerogel plate, and the cavity. 
	(b) Schematic description of the unit cell.}\vspace*{-5mm}
	\label{fig1}
\end{figure}

In this work, we make use of the efficient and unusual attenuating properties of silica aerogel plates to design subwavelength perfect sound absorbers. The analyzed system is depicted in Fig.~\ref{fig1}(a) and consists of a periodic repetition of the resonant building units illustrated in Fig.~\ref{fig1}(b). These units are made of a slit loaded by a clamped aerogel plate backed by a closed cavity. The perfect absorption of this system is comprehensively analyzed both theoretically and experimentally. In a first stage, we model the system using the Transfer Matrix Method (TMM) accounting for the contribution of the losses to the problem, i.e., the viscothermal losses from the slit and cavity and the viscoelastic losses from the aerogel plate. In a second stage, we analyze the impedance matching condition, also known as critical coupling condition, which is obtained once the inherent losses exactly compensates the leakage of the system\cite{Romero_2016a}. The Argand diagram of the reflection coefficient \cite{brekhovskikh2012acoustics} is further employed to evaluate either the lack or excess of inherent losses in the system, thus providing important information on the impedance matching condition. The Argand diagram is revealed as a universal and powerful tool to design perfect absorbers.
\par
We consider a slotted panel of thickness $L$, whose slits, of height $h_s$, are loaded by a circular clamped aerogel plate of radius $r_m$ and thickness $h_m$ backed by a cylindrical air cavity of the same radius and depth $l_c$. The clamped aerogel plate plays the central role, as it represents the main source of intrinsic losses and mainly governs the resonance of the system. Although the theoretical calculations are only carried out for the building block of size $L\times a\times a$ shown in Fig.~\ref{fig1}(b), the generalization to the case of $N$ resonator unit cell is straightforward.
%
%

For wavelengths $\lambda$ large enough compared to the thickness of the aerogel plate $h_m$ and neglecting the effects of rotary inertia and additional deflections caused by shear forces, the transverse plate displacement $w_m$ satisfies the Kirchhoff-Love wave equation \cite{Graff91}.
The plate can be described by $\rho$, the density and $D={Eh_m^3}/{12(1-\nu^2)}$, the bending stiffness. $E$ is the Young modulus and $\nu$ is the Poisson's ratio of the plate. 
Assuming an implicit time dependence $e^{i\omega t}$, with $\omega$ the angular frequency, the viscoelastic behavior of silica aerogel can be modeled via a complex Young modulus, $E=E_0(1+i\eta\omega)$, where $E_0$ and $\eta$ are the unrelaxed Young modulus and the loss factor respectively. 
In the subwavelength regime, the silica aerogel disk can be considered as a punctual resonant element located at $(x,y)=(L/2,a/2)$ (note that $h_m\ll\lambda_0$, where $\lambda_0$ is the wavelength in air). 
The acoustic impedance of the clamped circular cross-sectional plate thus takes the form \cite{Skvor91,Bongard10},
\begin{equation} \label{eq.Zp}
Z_p = \frac{-i\omega\rho h_m}{\pi r_m^2}\frac{ I_1(k r_m) J_0(k r_m) + J_1(k r_m)I_0(k r_m)}{I_1(k r_m)J_2(k r_m)-J_1(k r_m)I_2(k r_m)},
\end{equation}
where $J_n$ and $I_n$ are the of $n$-th order regular and modified Bessel's functions of the first kind respectively and the wavenumber in the plate satisfies $k^2=\omega\sqrt{\rho h_m/D}$. 
\par
Viscothermal losses also occurs in the narrow slits\cite{wardPRL2015} and in the cavity, also offering a useful degree of freedom to tune the losses of the system. 
Assuming only plane wave propagates in these channels, the viscothermal losses are modeled by effective parameters: complex and frequency-dependent wavenumbers $k_s=\omega\sqrt{\rho_s/\kappa_s}$ and  $k_c=\omega\sqrt{\rho_c/\kappa_c}$, and impedances $Z_s=\sqrt{\kappa_s\rho_s}/h_s a$ and $Z_c=\sqrt{\kappa_c\rho_c}/\pi r_m^2$ in the slit and in the cavity respectively. Note that we make use of the effective density $\rho_s$ and bulk modulus $\kappa_s$ of a slit \cite{stinson1991} for the slotted channel, while we make use of the effective density $\rho_c$ and bulk modulus $\kappa_c$ of a cylindrical duct\cite{stinson1991} for the cavity. 

\begin{figure*}[t]
	\centering
	\includegraphics[width=1\textwidth]{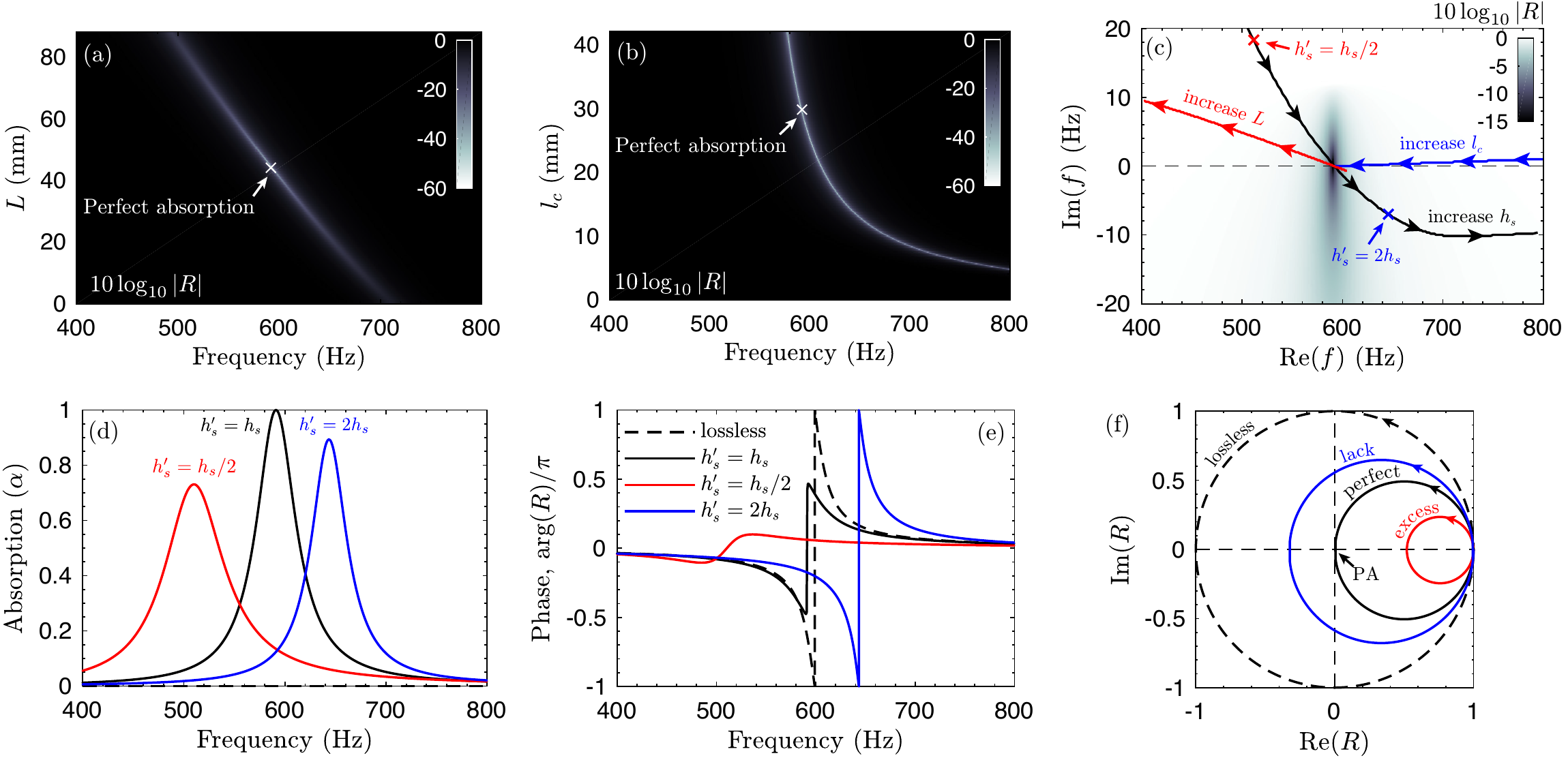}
	\caption{(color online) Reflection coefficient, in logarithmic scale, as a function of the frequency and (a) slit length, $L$, and (b) cavity depth, $l_c$. The rest of parameters are fixed to the optimal geometry (see main text). 
	(c) Representation of the reflection coefficient in the complex frequency plane for the sample with optimized parameters. 
	The lines show the trajectory of the zero when the corresponding geometrical parameter is modified. 
	(d)  and (e) show the absorption coefficient and the phase of the reflection coefficient respectively for three values of the slit thickness, $h^\prime_s$, $h_s$ (black line) being the optimum value. 
	(f) Argand diagram of the complex reflection coefficient from 0 to 1200 Hz, for the lossless structure (dashed circle), the case of perfect absorption (PA) geometry (black circle), the case $h_s'=h_s/2$, (red circle) and the case $h_s'=2h_s$ (blue circle). The small arrows indicate the trajectory from low to high frequencies.
	}\vspace*{-6mm}
	\label{fig2}
\end{figure*}

The scattering properties of the system are studied through the reflection coefficient $R$ obtained by TMM. 
We relate the sound pressure and normal particle velocity at the surface of the system, $[P_0,V_0] = [P(x), V_x(x)]_{x=0}$ to the ones at the end of the system, $[P_L,V_L] = [P(x), V_x(x)]_{x=L}$, by a transfer matrix as
\begin{equation}\label{eq.TMM}
\left[\begin{array}{c}
P_0 \\
V_0 \\
\end{array}\right] = {\bf T} \left[\begin{array}{c}
P_L \\
V_L \\
\end{array}\right],\,
\text{where}\quad
{\bf T}={\bf M}_{\Delta l}{\bf M}_S{\bf M}_{R}{\bf M}_S.
\end{equation}
In Eq.~(\ref{eq.TMM}), transfer matrix over half the slit length ${\bf M}_S$ reads as
\begin{equation}
{\bf M}_S=\left[\begin{array}{cc}
\cos\left(k_s{L}/{2}\right) & i Z_s\sin\left(k_s{L}/{2}\right) \\
{i}\sin\left(k_s{L}/{2}\right)/{Z_s} & \cos\left(k_s{L}/{2}\right) \\
\end{array}
\right],
\end{equation}
and ${\bf M}_{R}$ accounts for the local effect of the aerogel plate together with the back cavity as
\begin{equation}
{\bf M}_{R}=\left[\begin{array}{cc}
1 & 0 \\
{1}/{Z_{R}} & 1 \\
\end{array}
\right],
\end{equation}
where $Z_{R}=Z_p - i Z_c\cot(k_c l_c)$. The matrix ${\bf M}_{\Delta l}$ provides the radiation correction of the slit to the free space as
\begin{equation}
{\bf M}_{\Delta l}=\left[\begin{array}{cc}
1 & Z_{\Delta l} \\
0 & 1 
\end{array}
\right],
\end{equation}
where $Z_{\Delta l}=-i\omega\Delta l\rho_0/\phi_sa^2$, with $\phi_s=h_s/a$ the surface porosity of the metasurface, $\rho_0$ the air density and $\Delta l$ the end correction length that can be approximated as\cite{kergomard1987}
\begin{equation}
\Delta l=h_s\phi_s\sum_{n=1}^{\infty}\frac{\sin^2(n\pi\phi_s)}{\left(n\pi\phi_s\right)^3}.
\end{equation}
The surface impedance at $x=0$ can thus be directly obtained using Eq. (\ref{eq.TMM}) and considering the rigid backing condition ($V_L = 0$) as 
\begin{widetext}
	\begin{equation}\label{eq.ZT}
		Z_T = \frac{P_{0}}{V_{0}}=  \dfrac{Z_s (Z_{\Delta l}+Z_{R})+i \tan \left({k_s L}/{2}\right) \left[i Z_s Z_{R} \tan \left({k_s L}/{2}\right)+2 Z_{\Delta l} Z_{R}+Z_s^2\right]}{Z_s+ 2 iZ_{R} \tan \left({k_s L}/{2}\right)}.
	\end{equation}
\end{widetext}

Finally, we calculate the reflection and absorption coefficient using Eq.~(\ref{eq.ZT}) as
\begin{equation}\label{eq.Ref}
R=\frac{Z_T-Z_0}{Z_T+Z_0},\quad\text{and}\quad \alpha=1-|R|^2,
\end{equation}
where $Z_0$ is the impedance of the surrounding medium, i.e., the air\footnote{For the air medium at room temperature and ambient pressure $P_0 = 101325$ Pa, we used an adiabatic coefficient $\gamma = 1.4$, a density $\rho_0 = 1.213$ kg/m$^3$, a bulk modulus $\kappa_0 = \gamma P_0$, a Prandtl number $\mathrm{Pr} = 0.71$, a viscosity $\eta_0 = 1.839\times 10^{-5}$ Pa$\cdot$s, a sound speed $c_0 = \sqrt{\gamma P_0/\rho_0}$ m/s, and an acoustic impedance $Z_0 = \rho_0 c_0 / a^2$.}. For the aerogel plate, we used an unrelaxed Young modulus $E_0=197.92$ kPa and a loss factor $\eta=4.47\times10^{-6}$ Pa$\cdot$s, a density $\rho = 80$ kg/m$^3$ and a Poison's ratio $\nu = 0.12$, as characterized in Ref. [\onlinecite{Geslaina_2018}]. Aerogel plates of $r_m=19.5$ mm and $h_m=10.5$ mm were selected and $a=42$ mm was fixed by the width of the square cross-sectional impedance tube that was used for the experimental validation.
\par
The procedure begins with looking for the geometric parameters giving the most efficient absorption at the lowest resonance frequency. The geometric parameters of the system are thus optimized numerically using a sequential quadratic programming (SQP) method \cite{powell1978}. 
The following parameters were obtained: $L=44$ mm, $h_s=1.285$ mm, and $l_c=29.8$ mm. The highly efficient absorption peak also appears at 591.2 Hz and is associated with a reflection coefficient amplitude of $10\log_{10}|R| = -62$ dB. The corresponding perfectly absorbed wavelength is $\lambda/L=13.1$ times larger than he depth of the structure. This subwavelength feature is due to the slow sound properties induced by the presence of the slit loading resonators\cite{Groby2015,Groby2016,jimenez2016}. Effectively, the resonance frequency of the slit in the absence of these loading resonators is around 1900 Hz. \footnote{Note that the slow sound properties can be improved by using thinner aerogel plates, thereby allowing a strongly reduced ratio, e.g., $\lambda/L=30$ at $f=259$ Hz using $h_m = 1.32$ and $h_s= 0.95$ mm. Nevertheless, we are constrained by the available aerogel plates ($h_m=10.5$ mm) for the experimental validation.}

%
%
\begin{figure*}[t]
	\centering\vspace*{-2mm}
	\includegraphics[width=1\textwidth]{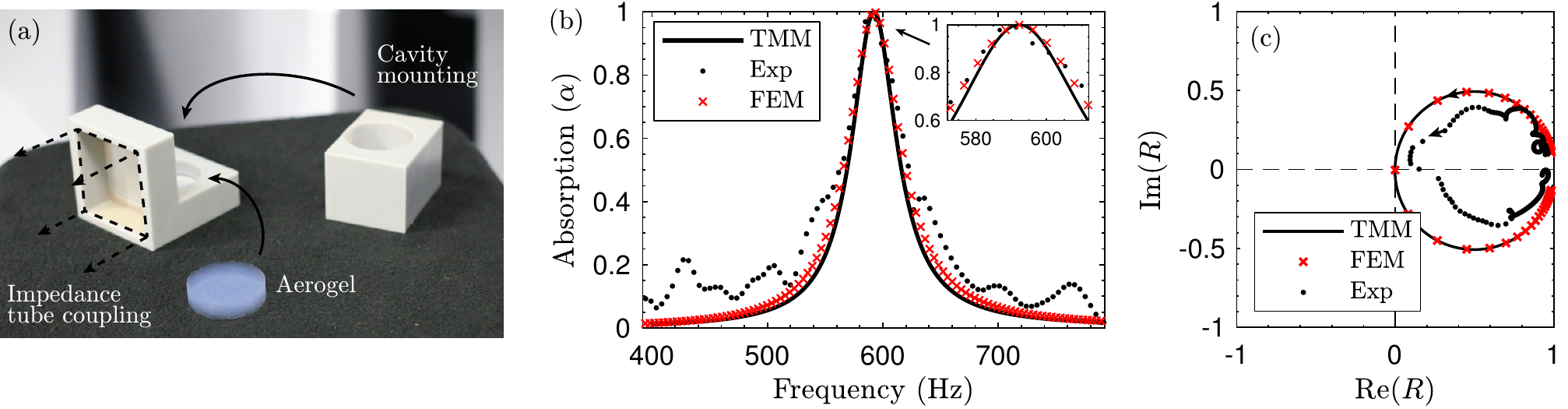}
	\caption{(color online) (a) Photograph of the experimental configuration. 
	(b) Absorption as a function of the frequency. 
	(c) Complex plane representation of the reflection coefficient. The small arrows indicate the trajectory from low to high frequencies.}\vspace*{-6mm}
	\label{fig3}
\end{figure*}
%

Figures~\ref{fig2}(a) and \ref{fig2}(b) show a parametric study of the system reflection coefficient around the optimal configuration. 
The reflection coefficient is significantly reduced when the parameters correspond to the optimal parameters (marked by white crosses). 
However, the balance between the inherent losses and the leakage of the system is difficult to identify by using this parametric analysis.

A first approach to ensure that these parameters led to perfect absorption of the sound energy consists in representing the reflection coefficient in the complex frequency plane, as shown in Fig.~\ref{fig2}(c). 
Using this representation, the locations of the zero/pole pairs of the reflection coefficient can be studied. 
In the lossless case, the zeros are complex conjugates of their corresponding poles, both appearing in the opposite half spaces of complex frequency plane (zeros in the lower half space and poles in the upper one with our time Fourier convention). 
However, the zeros follows a given trajectory towards the pole half space when losses are introduced in the system. Note that the losses are modified when modifying the system geometry. In this way, the trajectory of the lowest frequency zero is depicted Fig.~\ref{fig2}(c), when $h_s$, $l_c$ and $L$ are modified.  For a given set of geometric parameters, the trajectory of the zero crosses the real frequency axis, ensuring the balance of the leakage by the inherent losses, therefore providing the perfect absorption\cite{Romero_2016a}.
\par
The complex frequency plane also gives useful insights to design and tune open lossy resonant systems\cite{Romero_2016,jimenez2016,Jimenez2017}. Such system is characterized by its leakage and the inherent losses; the impedance matching condition corresponds to the critical coupling of the system, i.e., the perfect balance of the leakage by the inherent losses. On the one hand,  the intrinsic losses of the system are too large (too small) compared to the leakage of the system when the zero has (has not) already crossed the real axis, meaning that the absorption is not optimal as the impedance condition is not satisfied. 
These situations are illustrated Fig.~\ref{fig2}(d), where the absorption coefficient is depicted for different values of $h_s$. 
The red curve corresponds to a narrow slit (height $h_s'=h_s/2$) providing an excess of losses, while the blue curve corresponds to a wide slit (height $h_s'=2 h_s$) providing a lack of losses. 
The location of the corresponding zero in the complex frequency plane is marked with red and blue crosses in Fig.~\ref{fig2}(c). The perfect balance between the leakage by the losses is the situation depicted on the color map Fig.~\ref{fig2}(c); the zero of the reflection coefficient is exactly located on the real frequency axis. 
Therefore, we can conclude that the complex frequency plane is very useful to immediately identify if one particular configuration has a lack or excess of losses. 
This approach has been recently used to design absorbing materials ranging from porous media\cite{JimenezActa2018,jimenez2016broadband} to different kind of metamaterials\cite{JImenezAppSci2017,Jimenez2017,Jimenez2017b,Groby2016,Romero_2016}. However, the acoustic behavior of all systems cannot necessarily be assessed in the complex frequency plane. Effectively, numerical methods do not usually allow to calculate solutions for complex frequencies, and more importantly, experimental results are usually only provided for real frequencies. 

A useful approach to overcome this problem consists in analyzing the reflection coefficient $R=|R|e^{i\varphi}$, with $\varphi = \arctan [\textrm{Im}(R)/\textrm{Re}(R)]$, in the complex plane.
Figure ~\ref{fig2}(e) and (f) depict respectively the phase of the reflection coefficient and the corresponding Argand diagram from  400 to 800 Hz. The reflection coefficient is necessarily inscribed within the unitary circle, i.e., $|R|\leq1$. In the lossless case, the reflection coefficient follows the unitary circle counter-clockwise with increasing frequency starting from $\varphi=0$ at 0 Hz, as $R=e^{i\varphi}$. When losses are accounted for, the trajectory of $R$ is modified and follows an elliptical trajectory around the resonance, contained inside of the unitary circle and displaced along the real axis in the diagram. On the one hand, the reflection coefficient describes a loop that does not encompass the origin if the losses exceed the optimal ones, e.g. the red ellipse Figure ~\ref{fig2}(f) calculated for $h_s'=h_s/2$. On the other hand, the ellipse encompasses the origin if the losses lacks, e.g. the blue ellipse Figure ~\ref{fig2}(f) calculated for $h_s'=2h_s$. 
Finally, the ellipse must pass through the origin, i.e., $R=0$, when perfect absorption is reached, e.g. the black ellipse Figure ~\ref{fig2}(f) calculated for $h_s'=h_s$. 
In this situation, the impedance matching condition is satisfied. 
\par
The designed optimal structure was validated experimentally in a square cross-sectional impedance tube. 
The circular aerogel plate was cut by laser cutting and then inserted in a 3D-printed support manufactured by stereolithography (Form 2, Formlabs, UK). 
In addition, full wave numerical simulations by finite element method (FEM) were performed. 
For the FEM simulations, the plate was modeled as an elastic bulk plate of thickness $h_m$ considering a Kelvin-Voigt viscoelastic model and viscothermal losses were accounted for in the ducts using effective parameters as previously introduced for TMM calculations. 
Figure \ref{fig3}(a) shows the 3D printed system together with the aerogel plate before assembling. 
The measured absorption is shown in Fig.~\ref{fig3}(b). A good agreement is observed between the measurements, FEM simulations and TMM predictions. 
The ripples observed in the experimental data are probably due to non-symmetrical errors during the manufacturing of the circular plate, as well as to the fact that the plate is not perfectly clamped.
Finally, Fig.~\ref{fig3}(c) presents the Argand diagram of the reflection coefficient measured and calculated with the TMM. Both curves passe through the origin at a specific frequency that corresponds to the one at which the system is impedance matched. 

In summary, we have designed and manufactured a resonant building block made of cavity backed aerogel clamped plates that is suitable and efficient for perfect sound absorbing panel. 
The experimental data agree with those predicted by both the one-dimensional TMM model and the FEM simulations. 
We have presented a universal methodology based on the complex representation of the reflection coefficient, i.e., its Argand diagram, to identify the lack or the excess of losses in the system. 
This tool can be used further to design complex metasurfaces for perfect sound absorption when the system cannot be evaluated at complex frequencies, thus helping in the rapid design of novel and efficient absorbing metamaterials. 

\begin{acknowledgments}
This work has been funded by the RFI \textit{Le Mans Acoustique}, R\'egion Pays de la Loire. N.J. acknowledges financial support from Generalitat Valenciana through grant APOSTD/2017/042. 
J.-P.G and V.R.G. gratefully acknowledge the ANR-RGC METARoom (ANR-18-CE08-0021) project and the HYPERMETA  project funded under the program \'Etoiles Montantes of the R\'egion Pays de la Loire.
J.S-D. acknowledges the support of the Ministerio de Econom\'{\i}a y Competitividad of the Spanish government, and the European Union FEDER through project TEC2014-53088-C3-1-R.
\end{acknowledgments}



%

\end{document}